\documentclass[runningheads]{llncs}
\usepackage{graphicx}
\begin{document}
\title{SARS-CoV-2 Coronavirus Data Compression Benchmark}
\author{Innar Liiv}
\authorrunning{I. Liiv}
\institute{Tallinn University of Technology, 15A Akadeemia Rd, 12618 Tallinn, Estonia
\email{innar.liiv@taltech.ee}}
\maketitle
\begin{abstract}
This paper introduces a lossless data compression competition that benchmarks solutions (computer programs) by the compressed size of the 44,981 concatenated SARS-CoV-2 sequences, with a total uncompressed size of 1,339,868,341 bytes. The data, downloaded on 13 December 2020, from the severe acute respiratory syndrome coronavirus 2 data hub of ncbi.nlm.nih.gov is presented in FASTA and 2Bit format. The aim of this competition is to encourage multidisciplinary research to find the shortest lossless description for the sequences and to demonstrate that data compression can serve as an objective and repeatable measure to align scientific breakthroughs across disciplines. The shortest description of the data is the best model; therefore, further reducing the size of this description requires a fundamental understanding of the underlying context and data. This paper presents preliminary results with multiple well-known compression algorithms for baseline measurements, and insights regarding promising research avenues. The competition's progress will be reported at \url{https://coronavirus.innar.com}, and the benchmark is open for all to participate and contribute.
\end{abstract}
\section{Motivation}
Marvin Minsky considered Kolmogorov, Chaitin, and Solomonoff's algorithmic information theory ``the most important discovery since Gödel'' and conjectured that ``practical approximations to [their theory]\ldots would make better predictions than anything we have today'' \cite{minsky}.

This competition is intended to encourage multidisciplinary research, in the spirit of Kolmogorov, Chaitin, and Solomonoff's theory, to develop the shortest lossless description for the sequences of SARS-CoV-2. A successful result will serve as a demonstration that data compression can offer an objective and repeatable measure to align scientific breakthroughs across disciplines. The shortest description of a dataset is the best model. Further compression of the sequences of SARS-CoV-2  will require a fundamental understanding of the data and its context.

\section{Related Work}
The main theoretical underpinnings of this benchmark are the practical approximations to  Kolmogorov--Chaitin--Solomonoff complexity by Li and Vitanyi \cite{li2008introduction} and the minimum description length principle \cite{rissanen1978modeling,grunwald2007minimum}.

Kolmogorov complexity is the length of the shortest effective description of an object \cite{kolmogorov1965three}. Therefore, the idealistic goal of the SARS-CoV-2 coronavirus data compression benchmark is to find the Kolmogorov complexity of SARS-CoV-2 \cite{wu2020new} sequences. Since doing so requires an infinite amount of work, as a practical approximation, the smallest archive plus the decompressor is considered a computable proxy. Matt Mahoney has written an inspiring and excellent rationale for a large text compression benchmark \cite{mahoney2009rationale} with an extended discussion about the connections between intelligence and compression.

Several lossless compression benchmarks have been proposed over the years \cite{calgary,hutter2006}, the most well-known by Marcus Hutter, who offered €500,000 as a challenge prize \cite{hutter2006}. The compression of genetic sequences, as a specific niche of data compression research, has been a popular topic for more than 25 years \cite{grumbach1993compression,grumbach1994new}. An interested reader is referred to two comprehensive surveys about data compression methods for biological sequences \cite{bonfield2013compression,hosseini2016survey}. Kryukov et al. have recently presented a comprehensive evaluation of reference-free compressors for FASTA-formatted sequences \cite{kryukov2020sequence} and developed a sequence compression benchmark database.

De Maio et al. have identified several oddities specific to SARS-CoV-2 sequencing data, which may ``arise from specific combinations of sample preparation, sequencing technology, and consensus calling approaches'' \cite{de2020issues}. Such aspects, and other systematic errors typical to sequence data \cite{meacham2011}, can support the design of a specific compression strategy.

\section{The Task}
Losslessly compress the 1.25GB file \textit{coronavirus.fasta} \cite{innar} or its 2bit representation equivalent \textit{coronavirus.2bit} (0.31 GB) \cite{innar} to less than 1,238,330 bytes (the current smallest compressed size of the dataset, including the decompressor).

\section{The Data}
The data is presented in FASTA and 2Bit (UCSC-twobit \cite{karolchik2003ucsc}) format, consisting of 44,981 concatenated SARS-CoV-2 sequences with a total uncompressed size of 1,339,868,341 bytes \cite{innar}. It was downloaded on 13 December 2020 from the severe acute respiratory syndrome coronavirus 2 data hub of ncbi.nlm.nih.gov \cite{ncbi}. Each participant can choose which file to use---that is, the compressor does not have to work on both datasets.
\clearpage
\section{Setting the Scene}
The challenge is to compress 44,981 concatenated SARS-CoV-2 sequences. To provide a slightly simpler example, more susceptible to manual observation, the compression results for one sequence (reference sequence NC\_045512 \cite{wu2020new}) are presented in Table \ref{table1}.

\begin{table}
\caption{Compression results of the the reference sequence NC\_045512 (fewer bytes is better)}
\label{table1}
\begin{tabular}{| l | l | l | l | l |}
\hline
Bytes & File & Format & Compressor & Parameters \\
\hline
7233 & NC\_045512 & FASTA & cmix \cite{cmix} & ~ \\
7277 & NC\_045512 & FASTA & paq8l \cite{paq8l} & -8 \\
7277 & NC\_045512 & FASTA & GeCo3 \cite{silva2020efficient} & -l 1 -lr 0.06 -hs 8 \\
7308 & NC\_045512 & 2Bit & cmix \cite{cmix} & ~ \\
7337 & NC\_045512 & 2Bit & brotli \cite{brotli} & -q 10 \\
7346 & NC\_045512 & 2Bit & paq8l \cite{paq8l} & -8 \\
7355 & NC\_045512 & 2Bit & zstd \cite{zstandard} & -19 \\
7369 & NC\_045512 & 2Bit & bcm \cite{bcm} & -9 \\
7376 & NC\_045512 & 2Bit & gzip \cite{salomon2004data} & -9 \\
7508 & NC\_045512 & 2Bit & xz \cite{collin2010xz} & -9 \\
7517 & NC\_045512 & 2Bit & zip \cite{salomon2004data} & -9 \\
7524 & NC\_045512 & 2Bit & \textit{Uncompressed} & ~ \\
7545 & NC\_045512 & 2Bit & rar \cite{rar} & m5 \\
7802 & NC\_045512 & FASTA & bcm \cite{bcm} & -9 \\
7868 & NC\_045512 & 2Bit & bzip2 \cite{bzip2} & -9 \\
8399 & NC\_045512 & FASTA & brotli \cite{brotli} & -q 11 \\
8519 & NC\_045512 & FASTA & zstd \cite{zstandard} & -19 \\
8801 & NC\_045512 & FASTA & bzip2 \cite{bzip2} & -9 \\
9000 & NC\_045512 & FASTA & xz \cite{collin2010xz} & -9 \\
9598 & NC\_045512 & FASTA & gzip \cite{salomon2004data} & -9 \\
9623 & NC\_045512 & FASTA & rar \cite{rar} & m5 \\
9738 & NC\_045512 & FASTA & zip \cite{salomon2004data} & -9 \\
30416 & NC\_045512 & FASTA & \textit{Uncompressed} & ~ \\
\hline
\end{tabular}
\end{table}

\clearpage
\section{Preliminary Results}
Table \ref{table2} presents the current compression results for for 44,981 SARS-CoV-2 sequences (with a total uncompressed size of 1,339,868,341 bytes) sorted by the number of bytes (with fewer bytes meaning better compressibility), acting as the baseline measurement for the challenge. The bytes column in Table \ref{table2} did not include the size of the decompressor, which will be considered in the final benchmark. Considering the total size of the compressed archive and the decompressor (instead of just considering the compressed archive), the PAQ8L compressor \cite{paq8l} by Matt Mahoney performed the best, with the best results achieved using the 2Bit format of the dataset. The resulting compressed archive for PAQ8L, including the compressed decompression executable, has a total size of 1,238,330 (1,207,839+30,491) bytes. The CMIX compressor \cite{cmix} by Byron Knoll resulted a smaller compressed archive (988,958), but the total size, including the compressed decompressor, is 1,282,852 (988,958+293,894).

\begin{table}
\caption{Compression results for 44,981 SARS-CoV-2 sequences (fewer bytes is better)}
\label{table2}
\begin{tabular}{| l | l | l | l | l |}
\hline
Bytes & File & Format & Compressor & Parameters \\
\hline
988,958 & Coronavirus & 2Bit & cmix \cite{cmix} & ~ \\
1,207,839 & Coronavirus & 2Bit & paq8l \cite{paq8l} & -8 \\
1,425,590 & Coronavirus & FASTA & paq8l \cite{paq8l} & -8 \\
1,985,384 & Coronavirus & 2Bit & xz \cite{collin2010xz} & -9 \\
2,022,796 & Coronavirus & FASTA & xz \cite{collin2010xz} & -9 \\
2,043,140 & Coronavirus & 2Bit & bcm \cite{bcm} & -9 \\
2,044,664 & Coronavirus & 2Bit & rar \cite{rar} & m5 \\
2,050,285 & Coronavirus & FASTA & GeCo3 \cite{silva2020efficient} & -l 1 -lr 0.06 -hs 8 \\
2,367,487 & Coronavirus & 2Bit & zstd \cite{zstandard} & -19 \\
2,728,490 & Coronavirus & FASTA & bcm \cite{bcm} \cite{bcm} & -9 \\
2,871,864 & Coronavirus & 2Bit & brotli \cite{brotli} & -q 10 \\
2,871,864 & Coronavirus & FASTA & brotli \cite{brotli} & -q 10 \\
4,217,341 & Coronavirus & FASTA & zstd \cite{zstandard} & -19 \\
5,924,805 & Coronavirus & FASTA & rar \cite{rar} & m5 \\
67,575,178 & Coronavirus & 2Bit & gzip \cite{salomon2004data} & -9 \\
67,575,325 & Coronavirus & 2Bit & zip \cite{salomon2004data} & -9 \\
75,530,790 & Coronavirus & 2Bit & bzip2 \cite{bzip2} & -9 \\
75,530,790 & Coronavirus & FASTA & bzip2 \cite{bzip2} & -9 \\
77,356,405 & Coronavirus & FASTA & gzip \cite{salomon2004data} & -9 \\
77,356,550 & Coronavirus & FASTA & zip \cite{salomon2004data} & -9 \\
332,133,731 & Coronavirus & 2Bit & Uncompressed & ~ \\
1,339,868,341 & Coronavirus & FASTA & Uncompressed & ~ \\
\hline
\end{tabular}
\end{table}
\clearpage
\section{Conclusions}
The sequences of the SARS-CoV-2 coronavirus are compressible. Further compression will require a mix of novel and creative approaches: moving beyond the state of the art of data compression or understanding the patterns and relationships within parts of sequences and between sequences.

The SARS-CoV-2 coronavirus data compression benchmark has a vital multidisciplinary aspect: the objective and repeatable measure in this challenge can help to align scientific breakthroughs across disciplines. At the end of the day, different theories and models to understand the coronavirus are measurable through the shortest description of the dataset.

In addition, the scientific momentum around and attention paid to SARS-CoV-2 can be applied to support breakthroughs by the data compression community and advance the state of the art of compression. The techniques used for improving the compression of SARS-CoV-2 datasets can feed back to better understanding the underlying mechanisms of the coronavirus. 
\bibliographystyle{splncs04}
\bibliography{references}
\end{document}